\documentclass[sigconf]{acmart}
\newcommand{\SYSTEM}{CiteSee}

\newcommand{\jb}[1]{\textcolor{red}{[JB: #1]}}
\newcommand{\jc}[1]{\textcolor{blue}{[JC: #1]}}
\newcommand{\ah}[1]{\textcolor{brown}{[AH: #1]}}
\newcommand{\kl}[1]{\textcolor{orange}{[KL: #1]}}
\newcommand{\az}[1]{\textcolor{orange}{[AZ: #1]}}
\newcommand{\dd}[1]{\textcolor{violet}{[DD: #1]}}
\newcommand{\dw}[1]{\textcolor{violet}{[DanW: #1]}}
\newcommand{\update}[1]{\textcolor{blue}{#1}}
\newcommand{\upupdate}[1]{\textcolor{brown}{#1}}

\renewcommand{\jb}[1]{}
\renewcommand{\jc}[1]{}
\renewcommand{\update}[1]{#1}
\renewcommand{\upupdate}[1]{#1}
\renewcommand{\ah}[1]{}
\renewcommand{\kl}[1]{}
\renewcommand{\az}[1]{}
\renewcommand{\dd}[1]{}
\renewcommand{\dw}[1]{}

\newcommand{\new}[1]{\textcolor{blue}{#1}}
\renewcommand{\new}[1]{#1}

\AtBeginDocument{%
  \providecommand\BibTeX{{%
    \normalfont B\kern-0.5em{\scshape i\kern-0.25em b}\kern-0.8em\TeX}}}

\setcopyright{acmcopyright}
\copyrightyear{2018}
\acmYear{2018}
\acmDOI{XXXXXXX.XXXXXXX}

\acmJournal{TOG}
\acmVolume{37}
\acmNumber{4}
\acmArticle{111}
\acmMonth{8}

\copyrightyear{2023}
\acmYear{2023}
\setcopyright{acmlicensed}\acmConference[CHI '23]{Proceedings of the 2023 CHI Conference on Human Factors in Computing Systems}{April 23--28, 2023}{Hamburg, Germany}
\acmBooktitle{Proceedings of the 2023 CHI Conference on Human Factors in Computing Systems (CHI '23), April 23--28, 2023, Hamburg, Germany}
\acmPrice{15.00}
\acmDOI{10.1145/3544548.3580847}
\acmISBN{978-1-4503-9421-5/23/04}

\begin{document}

\title[{\SYSTEM}: Augmenting Citations in Scientific Papers with Personalized Context]{{\SYSTEM}: Augmenting Citations in Scientific Papers with Persistent and Personalized Historical Context}

\author{Joseph Chee Chang}
\email{josephc@allenai.org}
\affiliation{%
  \institution{Allen Institute for AI}
  \city{Seattle}
  \state{WA}
  \country{USA}
}

\author{Amy X. Zhang}
\email{axz@cs.uw.edu}
\affiliation{%
  \institution{University of Washington}
  \city{Seattle}
  \state{WA}
  \country{USA}
}

\author{Jonathan Bragg}
\email{jbragg@allenai.org}
\affiliation{%
  \institution{Allen Institute for AI}
  \city{Seattle}
  \state{WA}
  \country{USA}
}

\author{Andrew Head}
\email{head@seas.upenn.edu}
\affiliation{%
  \institution{University of Pennsylvania}
  \city{Philadelphia}
  \state{PA}
  \country{USA}
}

\author{Kyle Lo}
\email{kylel@allenai.org}
\author{Doug Downey}
\email{dougd@allenai.org}
\affiliation{%
  \institution{Allen Institute for AI}
  \city{Seattle}
  \state{WA}
  \country{USA}
}


\author{Daniel S. Weld}
\email{danw@allenai.org}
\affiliation{%
  \institution{Allen Institute for AI \&\\University of Washington}
  \city{Seattle}
  \state{WA}
  \country{USA}
}

\renewcommand{\shortauthors}{Chang et al.}

\begin{abstract}
\update{When reading a scholarly article, inline citations help researchers contextualize the current article and discover relevant prior work. However, it can be challenging to prioritize and make sense of the hundreds of citations encountered during literature reviews. This paper introduces {\SYSTEM}, a paper reading tool that leverages a user's publishing, reading, and saving activities to provide personalized visual augmentations and context around citations. First, {\SYSTEM} connects the current paper to familiar contexts by surfacing known citations a user had cited or opened. Second, {\SYSTEM} helps users prioritize their exploration by highlighting relevant but unknown citations based on saving and reading history. We conducted a lab study that suggests {\SYSTEM} is significantly more effective for paper discovery than three baselines. A field deployment study shows {\SYSTEM} helps participants keep track of their explorations and leads to better situational awareness and increased paper discovery via inline citation when conducting real-world literature reviews.
}

\end{abstract}

\ccsdesc[500]{Human-centered computing~Graphical user interfaces}

\keywords{reading interfaces, scientific papers, personalization}

\maketitle

\begin{figure}[H]
\centering
	\includegraphics[width=0.8\linewidth]{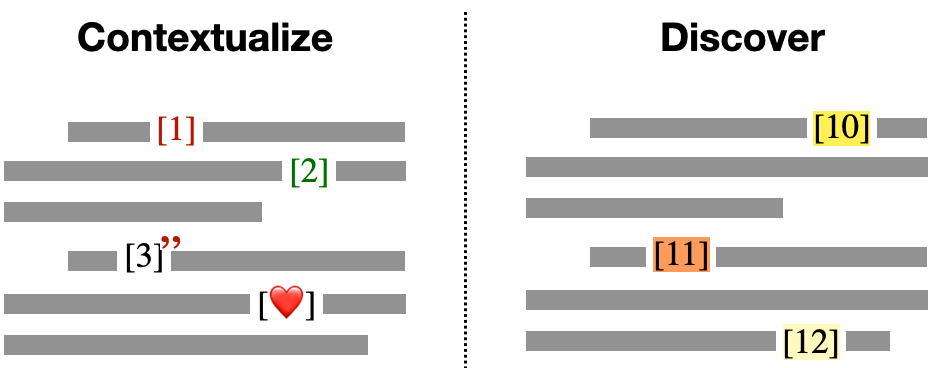}
	\caption{{\SYSTEM} augments inline citations to known papers to help contextualize the current paper. This includes saved (1, red) and visited papers (2, green), papers previously cited by current user (3''), and their own publications ($\heartsuit$). {\SYSTEM} also highlights citations to unknown papers (10-12) to help discover important prior work based a user's engagements on their citing papers.}
	\label{fig:teaser}
 \Description{Left: Titled "contextualize" at the top, followed by a paragraph figure with 4 inline citations. Citation 1 is red, 2 is green, 3 is overlaid with a red quotation mark, and 4 is overlaid with a red heart emoji. Right: Titled "Discover" at the top, followed by a paragraph figure with 3 inline citations. Citation 12 is highlighted in light yellow, and citation 10 is highlighted in a more saturated yellow. Citation 13 is highlighted in a very saturated yellow-orange color.}
\end{figure}

\begin{figure*}[t]
    \centering
    \begin{minipage}{0.46\textwidth}
        \centering
        \fbox{\includegraphics[width=1\textwidth]{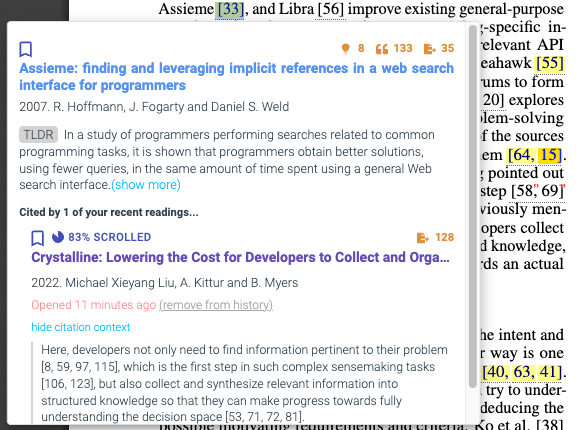}}
    \end{minipage}\hfill
    \begin{minipage}{0.524\textwidth}
        \centering
        \fbox{\includegraphics[width=1\textwidth]{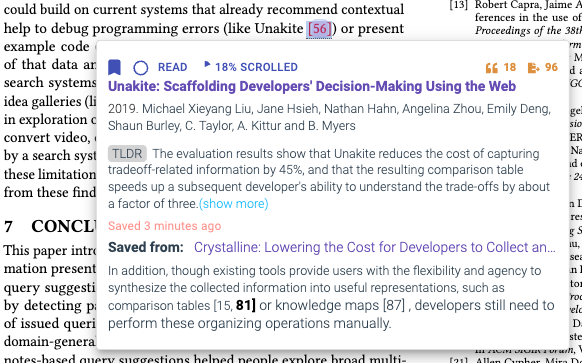}}
    \end{minipage}
    \caption{
    [Left] To help users discover important prior work, unexplored citations are highlighted in different shades of yellow to indicate their potential relevance to the user. [Right] To help users keep track of which citations were already explored and to draw connections between familiar papers to the current paper, inline citations to familiar papers (e.g., saved) are rendered in red. [Both]  To see personalized context around inline citations, users can click on a citation to see its Paper Card with personalized context such as citing sentences from recently read papers or the citing sentence where the cited paper was saved. 
    } 
    \Description{Two side-by-side screenshots of a partial screen. Left: A popup paper card on top of a research paper. The corresponding citation [33] is highlighted and yellow and is the corresponding inline citation to the popup card. The content of the card is as follows: The main area has the title, publication year, authors, and abstract of the citing paper. The upper left corner has a bookmark button, and the upper right corner contains the citation and references count of the citing paper. The bottom half of the card contained a list of other paper titles from the user's reading history as well as the citing sentences from these papers. Right: A similar screenshot, in this one, the paper card contained a previously saved paper. Citation [56] is red showing that it is previously saved. Besides the same information as the paper card shown on the left, this paper card also contained "Saved from:" a paper title and a citing sentence.}
    \label{fig:paper_cards}
\end{figure*}

\section{Introduction}

Science builds on the past work of others. Researchers draw from prior work to synthesize existing knowledge, identify research opportunities, and find inspirations for future research. One of the fundamental ways researchers explore and learn from the literature is by reading scientific papers. This not only provides them insights into individual prior work, but the related work sections also allows scholars to discover and draw connections to additional relevant papers via inline citations \cite{King2009ScholarlyJI}. This process allows researchers to contextualize the paper they are reading within cited work, become aware of research threads that influenced the current paper, and discover other important and relevant papers to further their literature reviews \cite{Tenopir2008ElectronicJA,Evans2008ElectronicPA,King2009ScholarlyJI}.
\new{Inline citations are a key resource for discovering papers. The behavior of following multiple levels of inline citations, sometimes referred to as \emph{chaining} or \emph{footnote chasing}, has been observed across many scholar groups such as sociology, computer science, and economics (summarized in \cite{palmer2009scholarly}). More specifically, one survey study estimated that inline citations accounted for around one in five (21\%) of paper discoveries during research \cite{King2009ScholarlyJI}.}

\upupdate{While inline citations are useful for discovering literature, it is often difficult to prioritize which citations to pay attention to in the middle of a reading task. One challenge is that even though there is some relationship between all inline citations and the citing paper, only a subset of them will be relevant to the reader's interests at the time of reading.} \update{This is especially challenging during literature reviews, where users need to read and skim many papers, each of which may contain dozens or hundreds of inline citations. For example, a user interested in learning about \emph{text analysis techniques} reading a paper about \emph{sentiment analysis on customer reviews} might be interested in inline citations to prior work in \emph{natural language processing} but not \emph{e-commerce marketing}.}

\upupdate{Recently, research systems have been developed to help readers discover papers. HCI researchers have designed numerous standalone interactive paper discovery tools to support exploration of large corpora of papers (e.g.,~ \cite{He2019PaperPolesFA,Ponsard2016PaperQuestAV,Chau2011ApoloMS}). NLP researchers have developed technologies that analyze inline citations in a way that could be assistive to understanding those citations, for instance classifying their level of influence on the citing paper~\cite{Valenzuela2015IdentifyingMC} or predicting their intent (e.g., whether the citation informs the methods, background, or results)~\cite{Cohan2019StructuralSF}.}

\upupdate{What readers do not have, but could benefit from, are tools that provide in-situ support, within a paper, for the challenging task of understanding how citations relate to their own nuanced, evolving research interests and search history. Such an understanding of citations is necessary for deciding which of many citations are worth consulting. The purpose of this paper is to design and evaluate usable in-situ aids for prioritizing inline citations.}




\update{\upupdate{The key insight motivating our eventual design for citation prioritization aids arose from need-finding interviews (described in Section \ref{sec:needfinding}): participants wished for a tool that helped them keep an eye out for prior work that is cited by {\em multiple} papers they had read in a literature review.} To continue the scenario above, if a user noticed a paper cited from both a paper about \emph{aspect extraction on customer reviews} and another paper about \emph{sentiment analysis on news articles}, the cited paper was expected to be more relevant and salient to the reader's interest of \emph{text analysis techniques}.
However, keeping track of which papers are cited by multiple papers during a literature review is impractical in current reading tools: papers use opaque identifiers for citations, like reference numbers or author-year abbreviations that differ across papers. Current reading tools do not keep track of which citations a reader has seen before (a basic affordance that sees widespread use in web browsers, which render hyperlinks in purple color when they have already been visited).  Even if a reader does recognize a citation that they have seen in another, they likely will not be able to recall the context from which it was cited in other papers (e.g., which sections and the citing sentences), making it difficult to assess their importance and relevance across their corpus. These factors led participants in our preliminary interviews (described in a later section) to point a concern of ``missing out'' on prior work that is well-known and frequently cited by other researchers working on similar topics.}

In this paper, we introduce and explore the idea of a personalized paper reading experience that augments citations in a reading tool based on their connections to the current user. We developed a Chrome-extension PDF reader for scientific papers called {\SYSTEM}. Leveraging a user's paper library,  publication record, and reading history, {\SYSTEM} visually augments scientific papers to help users keep track of citations to known papers and prioritize their exploration to citations to unknown prior work that were likely relevant to their literature review topics (Figure~\ref{fig:types}).  \new{One key motivation here is that a user's publications and paper libraries can potentially represent their longer-term research interests, and their recent paper reading history can potentially represent their fluid and shorter-term research interests, such as during literature reviews for new projects.} In addition to visually augmenting inline citations, to help users better make sense of the cited papers, {\SYSTEM} keeps track of a consistent and personalized context of how different papers connect to the user's previous activities, for example, reminding users of the context of how they discovered different papers saved in their library or how an inline citation was described by other papers in their reading history (Figure~\ref{fig:paper_cards}).
The final design of {\SYSTEM} was driven by need-finding interviews with five researcher participants with varying research experiences (described in a later section), as well as several months of internal testing, design, and evaluation by the research team. \upupdate{The primary design challenge we addressed was to develop in-situ indicators that were simultaneously \emph{deeply informative} about the contexts where a citation has been encountered before, while also being \emph{subtle}, integrating into a paper reading experience without distracting or overwhelming the reader.} This paper contributes:

\begin{enumerate}

\item A prototype scientific paper reading tool, {\SYSTEM}. \upupdate{While prior work either analyzes inline citations in a non-personalized way \cite{Valenzuela2015IdentifyingMC,Cohan2019StructuralSF} or only support personalized paper discovery independent of reading \cite{Ponsard2016PaperQuestAV,He2019PaperPolesFA,Chau2011ApoloMS}, {\SYSTEM} explores the idea of a personalized reading experience focused on helping users make sense of inline citations and prioritize which citations to further consult during reading. }

\item Mechanisms for augmenting inline citations that have connections to a user's previous activities and providing a consistent and personalized historic context to help users discover, save, and keep track of important prior work during literature reviews.





\item A controlled lab study (N=10) focusing on paper discovery during reading which shows our simple highlighting strategy was significantly more effective than three baselines, including one that utilizes a more sophisticated semantic embedding technique. 

\item \upupdate{A field deployment study (N=6) with real-world literature review tasks which offers qualitative insights of how {\SYSTEM} helped participants prioritize and keep track of explorations with results suggest a 2.7x increase in paper discovery rate via inline citations compared to previously reported numbers that were based on self-reporting \cite{King2009ScholarlyJI}.}

\end{enumerate}

\begin{figure*}[t]
    \centering
    \fbox{\includegraphics[width=1\textwidth]{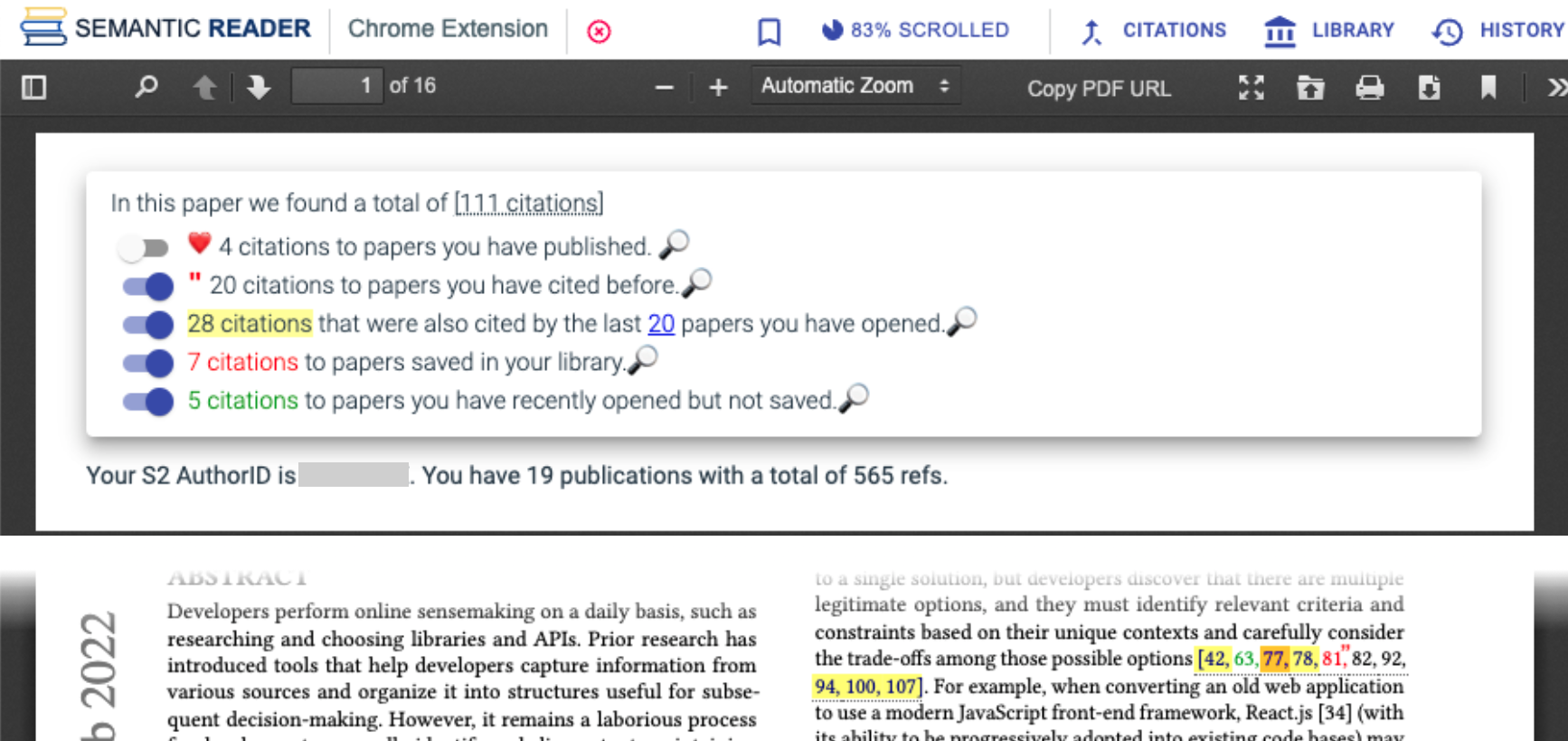}}
    \caption{{\SYSTEM} augments inline citations based on a user's reading history and paper library. [Top] An overview page inserted by {\SYSTEM} shows the statistics of different augmentations in the current paper. Users can toggle different augmentation types to avoid distraction. Users could also see Paper Cards for augmented inline citations in list views. [Bottom] Inline citations are visually augmented based on their connections to the current user. For example, to help users keep track of citations to already explored papers, previously opened or saved papers were rendered in green or red, respectively; to help users discover relevant and unexplored papers, citations also cited by papers in their reading history is highlighted in different shades of yellow. See Figure~\ref{fig:types} for different types of visual augmentations supported by {\SYSTEM}.
    } 
    
    \label{fig:hero}
    \Description{ A wide screenshot spanning two columns. At the top is a menu bar, from left to right are the following buttons: a red delete button, a bookmark button, a scroll position indicator showing "83\% scrolled", a citations panel button, a library panel button, and a history panel button. Below is a summary page that showed the current paper contained 111 citations, 4 were published by the current user, 20 were cited by the user, 28 were also cited by the last 20 papers the user has recently opened, 7 were papers saved in the current user's library, and 5 were paper recently opened but not saved. Finally, a line that reads "Your S2 Author ID is XXXX. You have 19 publications with a total of 565 refs. Below the summary page is a partial page from the current paper pdf. A few of the inline citations were highlighted in different shades of yellow, one is rendered in green and another one is rendered in red.}
\end{figure*}

\section{RELATED WORK}


\new{To better support the literature review process in a scientific paper reader interface, we look to prior work in exploratory search \cite{Marchionini2006FromFT}, sensemaking \cite{Russell1993TheCS}, and information foraging \cite{Pirolli1995InformationFI} behavior for user models to guide the design of our system. For example, we assume the process of literature review to be exploratory in nature, where users initially might not have clear ideas about the information they are seeking but crystallize their goals as they explore and learn from the literature \cite{Marchionini2006FromFT}. During this sensemaking process, users can also develop their own schemas for organizing the literature, such as forming different categories (e.g., subdomains) of prior work as they read, guiding their subsequent exploration \cite{Russell1993TheCS}. Most importantly, instead of deep reading each article, users often skim \cite{SchemaladenPA,Hillesund2010DigitalRS,Nicholas2010ResearchersEU} and switch between large numbers of scientific papers (i.e., \emph{information patches}) to optimize their information foraging efficiency (e.g., the rate of discovering important information and prior work) \cite{Pirolli1995InformationFI}. For example, by augmenting inline citations with behavior traces (e.g., citation statements from previous readings), {\SYSTEM} can potentially enrich the \emph{information scants} \cite{Pirolli1995InformationFI} between papers  to help users to better prioritize which inline citations to follow when conducting literature reviews.}

\subsection{Scientific Paper Reading Interfaces}

Research in general active reading has pointed to issues users often face when reading to conduct knowledge work \cite{Bianchi2015DesigningAP,Tashman2011ActiveRA,Tashman2011LiquidTextAF}. For example, one major challenge that users often struggle with is cross-referencing within and between documents \cite{OHara1998RankXR,Marshall1999IntroducingAD,Tashman2011LiquidTextAF,Bianchi2015DesigningAP}. Similarly, early research on scientific paper reading interfaces has focused on better support for cross-referencing citations. Specifically, they used heuristics and machine learning techniques to identify inline citations in research papers and map them to items in the references sections. This mapping enabled interactions that helped users avoid the high interaction costs of scrolling to the references section to contextualize citations as they read. For example, \cite{Powley2009EnrichingAD} demonstrated an interaction where users could click on a citation in the text to receive its corresponding title and authors in a popup card. Beyond lowering interaction costs, recent work has also explored cross-referencing relevant content to help users better understand the current paper. For example, ScholarPhi allowed users to cross-reference between terminologies or math symbols and their definitions interactively \cite{Head2021AugmentingSP}; and a thread of work focused on allowing users to point to different parts of a figure or table to highlight corresponding texts in the paper \cite{Kong2014ExtractingRB,Badam2019ElasticDC,Kim2018FacilitatingDR}. In contrast, instead of focusing on within-document cross-referencing for a single paper, we focus on supporting users in cross-referencing between scholarly papers as they reencountered the same citations to better support literature reviews, allowing users to identify and follow important citations to explore different papers while keeping track of how the same citation was described across different papers the users have read.

In a system more closely related to our work, CiteRead is a  \emph{non-personalized}, paper-reading tool that annotates an opened paper with relevant information extracted from \emph{incoming citations} of follow-on papers, allowing users to explore how the current paper has influenced later research \cite{Rachatasumrit2022CiteReadIL}. \update{In contrast, our work focuses on augmenting \emph{outgoing citations} to relevant prior work when reading multiple papers to help users make sense of the overwhelming quantity of relevant prior work based on their specific, \emph{personal interests}. Unlike prior work above that does not consider users' interests, {\SYSTEM} generates personalized annotations by exploiting users' recent reading history to capture their fluid interests during literature reviews.}

\subsection{Paper Recommendation and Exploration}

Besides improving the reading experience, research has also devoted significant efforts to helping researchers discover interesting papers to read independent of the reading experience \cite{Bai2019ScientificPR}. One major thread of research in machine learning has focused on recommender systems that allow users to rate a set of seed papers to train an agent that can provide a list of recommendations based on paper contents \cite{Sugiyama2010ScholarlyPR,Philip2014ApplicationOC}, the citation graph \cite{Huang2002AGR,Gori2006ResearchPR,Xia2016ScientificAR}, or a combination of the two \cite{Wang2011CollaborativeTM,Cohan2020SpecterDR}. For example, Specter \cite{Cohan2020SpecterDR} can generate a list of paper recommendations by computing document-level embeddings for scientific papers based on a language model trained on paper titles, abstracts, and their references. In our evaluation, we used Specter as one of our baseline approaches for ranking and recommending inline citations in a reading environment.
In contrast to generating lists of paper recommendations, HCI researchers have also explored interactive visual interfaces for exploring citation graphs. For example, early work from the 90s allowed users to search and explore forward and backward citations of a seed paper in a 3D environment \cite{Mackinlay1995AnOU}. More recently, Apolo and PaperPole took a mixed-initiative approach that allowed users to explore and create visual topic clusters of citations around a seed paper and provided paper recommendations for further refining the structures \cite{Chau2011ApoloMS,He2019PaperPolesFA}.
\new{PaperQuest allowed users to specify sets of seed papers as ``Core Interests'' or ``To-reads,'' which contributed different weights for ranking their references as recommendations \cite{Ponsard2016PaperQuestAV}, but did not present evaluations to justify this technique. Threddy allowed users to collect clips from papers and provided an exploration interface showing additional papers relevant to the collected clips \cite{kang2022threddy}, but did not support triaging exiting references in the current paper a user is reading.
More fundamentally, prior systems described above focus on developing a separate bespoke interface and do not support paper discovery during reading which accounted for a significant proportion how scholars became aware of prior work \cite{King2009ScholarlyJI}.
In contrast, we design a personalized reading tool to improve users' current behavior of discovering prior work through highlighting inline citations in-situ in a reading interface \cite{tenopir2012,King2009ScholarlyJI}. To highlight the inline citations for paper discovery, our scoring techniques is inspired by \cite{Ponsard2016PaperQuestAV}, but instead of requiring users to explicitly specify a set of paper of interest, we
exploited a user's paper reading history and engagement with the different papers to carefully incorporating these signals into the visually-dense medium of the paper with appropriate controls, and evaluating them both in-lab and field deployment studies in comparison to three baseline approaches. 
}
\new{Finally, \citet{hyeonsu} explored generating explanations of emailed paper recommendations based on personalized social signals (i.e., a paper's relations to familiar authors). While both \cite{hyeonsu} and {\SYSTEM} aimed to provide personalized contexts around papers, we focused on providing a personalized historical context around inline citations in a reading environment (e.g., the previously read paper and citing sentence where an inline citation was saved from) to support the literature review process where users are likely more concerned with finding papers based on topic relevance instead of social signals.}



\section{Preliminary Interviews}
\label{sec:needfinding}

\new{In early phase of the project, we conducted preliminary interviews to better inform ourselves about how researchers make sense of inline citations as they read, and the common limitations and needs that arise during literature reviews. This is primarily to help us develop a set of design goals listed in Section 4.3 to motivate system designs.}
For this, we recruited five participants with varying research backgrounds and experiences: 1 industry research manager, 1 assistant professor, 2 PhD students, and 1 predoctoral researcher working on HCI, CV, or NLP research.

Before the interviews, we generated eight scientific paper reader interfaces mock-ups aimed to address different potential issues as design probes. Similar to a scheme used in \cite{Powley2009EnrichingAD,Head2021AugmentingSP}, all eight designs allowed users to click on an inline citations and bring up paper context card with the title and abstract of the cited paper. In addition, different designs highlighted inline citations using different strategies and showed different additional context in their Paper Cards (described below in the context of our findings.)
In the first 20 minutes of the interview, participants searched online for an interesting paper to read, and performed think-aloud focusing on the inline citations as they encountered them. The second half of the interviews consisted of walking through the design mock-ups for 40 minutes, where we probed how strongly they reacted to the issues each design aimed to address. \new{The interviews were recorded for analysis. The first author went through the recordings and used an open thematic analysis process to capture qualitative insights \cite{Boyatzis1998TransformingQI,Connelly2013GroundedT}. The rest of this section lists the common themes from the five interviews, describes the design probes when relevant, and formulates our design goals based on these findings.}

\subsection{Fear of Overlooking Important Citations}
\new{
While some participants recalled paying less attention when reading to understand a paper quickly (e.g., to figure out the accuracy of a machine learning model), all participants emphasized the importance of using inline citations during literature reviews to discover important prior work. The most common sentiment was a fear of missing out on important prior work when citations were overlooked.
More specifically, they described how not being aware of closely related prior work could have severe consequences, even when it was not highly cited globally -- for example, putting significant research effort into an approach that had already been explored or being unaware of a paper that ``everyone else working on this are citing.''
}

\new{
We showed two design probes that annotated inline citations to facilitate paper discovery. The two reader designs that automatically highlighted inline citations in the current paper that either had similar titles to papers saved in a user's paper library or were cited by other recently read papers, respectively.
Participants reacted more positively to the second idea. For the first idea, we found participants had concerns around the accuracy of measuring semantic similarity between papers and the lack of explanations. 
In addition, participants also pointed to how their existing folders often represent longer-term interests that might not correspond to their interests during literature reviews which can be fluid and shorter-term. In contrast, when responding to the second idea, many participants recalled experiences in the past when they were reading different papers and noticed citations to the same prior work, often leading to important discoveries. However, they also agreed that this requires them to examine the right inline citation across different papers by chance and was not a signal that they could consistently notice.
}


\vspace{-2mm}

\subsection{Progress Tracking and Loss of Context}
\new{
Participants described using different strategies to save papers to read or keep references. For example, queuing papers in browser tabs, copying and pasting paper titles to external documents, or maintaining libraries and folders. One user challenge here was keeping track of sufficient context around saved papers. Participants recalled revisiting a saved paper but not remembering why it was saved (or kept opened in a browser tab.) Participants saw potential in designs that tracked their exploration trails, such as search queries and citing sentences relevant to a paper, to help them remember why it was saved in the first place. 
}

\new{
At a high level, while different participants had varied levels of interest in designs that augmented the inline citations in different ways, all participants responded positively to the idea of having consistent annotation and context for the same citations when reading different papers (i.e., citations to the same papers are annotated the same way across reading different papers). For example, all participants responded positively to a simple design that rendered inline citations in different colors based on whether they were previously opened in the reader or saved to their libraries. More fundamentally, participants expressed how it is high cost to synthesize information across different documents about the same papers. For example, using a separate spreadsheet or word document to keep track of important papers and maintain persistent context around them, such as collecting citing sentences across different papers they had read. 
}





\subsection{DESIGN GOALS}

Based on the above, we formulated the following design goals for a novel scientific paper reading interface to support the following during literature reviews:

\begin{itemize}
\item \textbf{[D1]} Augment citations to unknown papers that are also cited by papers in a user's reading history to help users discover prior work relevant to their literature reviews.
\item \textbf{[D2]} Augment citations to known papers (such as previously visited or saved papers) to connect the current papers to familiar contexts, helping users understand whether a paper belongs to a pocket of literature they have already explored or not.
\item \textbf{[D3]} Help users better make sense of inline citations by keeping track of how a user interacted with different papers to present consistent and personalized contexts. For example, clicking on an inline citation allows users to see how the cited paper is discussed across different papers that the user has read in the past.


\end{itemize}

\section{SYSTEM DESIGN}

Motivated by the design goals and exploratory interviews, we developed a novel scientific paper reader called {\SYSTEM}. When using current scientific paper reading interfaces, to become aware of citations to papers a user has seen before, they would have to recognize the papers either by reading the sentences around the citations or by searching through paper titles in the references section. In contrast, {\SYSTEM} keeps track of a user’s reading history and paper library to visually augment inline citations both to papers already explored in the past and to important but unexplored papers (Figure~\ref{fig:types}).
One challenge here is that users might not be able to remember their past interactions with different papers even when their inline citations are augmented by the system. To support this, {\SYSTEM} allows users to click on an inline citation to see personalized contexts in a Paper Card (Figure~\ref{fig:paper_cards}), such as the last time the paper was opened or the citing sentences from across papers they have recently explored.
Together, these features provide a personalized reading and exploration experience by augmenting and providing consistent and personalized historic context around citations.



\subsection{Overview of Citation Augmentation Types}

\begin{figure}[t]
\centering
    \vspace{0mm}
	\includegraphics[width=1\linewidth]{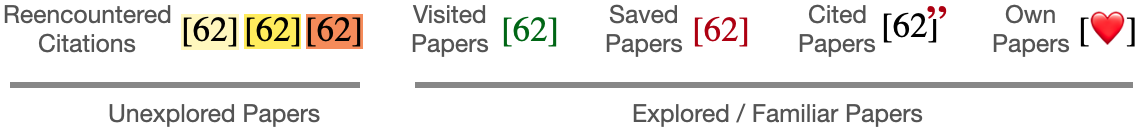}
	\caption{Overview of different visual augmentation types, with one category for citations to unexplored papers, and four categories for explored / familiar papers.}
	\label{fig:types}
 \Description{An overview of different visual augmentations. From left to right, common citation type with three examples with different yellow highlight shadings. Visited papers with 1 example rendered in green, saved papers with 1 example rendered in red, cited papers with 1 example with a red quotation mark overlaying the top right corner, and "own papers" with 1 example with a red heart emoji overlay the reference number. Under the common citations examples, there is an "unexplored papers" label, and under the other 4, there is an "Explored /familiar papers" label.}
\end{figure}

Similar to prior work in scientific paper reader interfaces \cite{Powley2009EnrichingAD,Head2021AugmentingSP}, {\SYSTEM} allows
users to interact with an inline citation by clicking query for information about a cited paper in a popup Paper Card (figure~\ref{fig:paper_cards}). The Paper Cards include the title, authors, publication year, abstract, abstract summary \cite{Cachola2020TLDRES}, and citation count of its corresponding inline citation. This allows users to make quick judgments about the inline citations without scrolling to the references section at the end of the paper. While prior systems focused on surfacing non-personalized context around citations, {\SYSTEM} also provides personalized context based on a user’s reading and publication history. For this, {\SYSTEM} visually augments the inline citations to indicate different ways the cited papers are connected to the current user. Figure~\ref{fig:types} shows an overview of how inline citations are visually augmented in our system as detailed below:

\begin{itemize}
\item \textbf{Reencountered Citations}: Citations that also appeared in other papers in a user’s reading history are highlighted in different shades of yellow to orange based on the user's engagement with the citing papers.
\item \textbf{Visited Papers}: Citations to papers previously opened by the user are rendered in green.\dw{isn't there some priority ordering? what about papers that are common but also visited? how do they appear?  Most saved papers will have been visited, but not all - eg i save papers to read later. if i've done both is it green or red?} \jc{added a snippet below}
\item \textbf{Saved Papers}: Citations to papers saved in a user’s library folders are rendered in red. 
\item \textbf{Cited Papers}: Papers that are cited by the user’s own publications are annotated with a red quotation mark at the upper right corner.
\item \textbf{Own Papers}: Heart emojis are rendered over the inline citations to the user's previous publications.
\end{itemize}

\noindent When multiple augmentations are applicable for the same inline citation a simple heuristic to prioritize them: If a paper was both visited and saved, we only apply the \emph{saved paper} augmentation. If a paper was previously published or cited by the user, we prevent \emph{reencountered citation} annotation from being applied to avoid over highlighting papers already known by the users for discovery.

In addition, when clicking on a inline citation, {\SYSTEM} provides personalized historic context in the form of Paper Cards (Figure~\ref{fig:paper_cards}), allowing users to see their past interaction related to the cited paper to help them recall a known paper or better make sense of an unexplored paper.
Below, we describe how {\SYSTEM} uses these core mechanisms to support different user needs when conducting literature reviews.
We first describe an example user scenario to ground our designs, and then unpack details of the various features and how they address the design goals above.

\subsection{Example User Scenario}

Consider an example where a researcher wants to learn more about \emph{automatic image captioning}. She starts searching on a scientific document search engine and opens a few papers from a search result that looks promising and published recently. As she reads the papers, she pays closer attention to the inline citations so that she can have good coverage of the important prior work in the area.
However, she wonders if she has skimmed through and missed some citations to important prior work, but there is also a diminishing return in spending time looking up the titles of inline citations in the references section since the chances of her encountering citations to papers she has already read increases as she explores more papers in the area.
More fundamentally, being new to the domain, it can sometimes be difficult for her to judge the importance of a citation based on how it was described in the current paper. For each citation she was unsure about, she could search online to check its citation count, abstract, and cross-reference to see how it was described in other papers that cited it. However, it would be too costly and disruptive to go through this process for each inline citation that might be useful.

Feeling overwhelmed, she switches to the {\SYSTEM} reader and opens up the same papers (Figure~\ref{fig:hero}), and can immediately see which inline citations were already opened in her browser tabs. After reading a couple of papers in {\SYSTEM}, she started to notices the reader highlighted citations based on her reading history. She can now see inline citations to papers she has already opened rendered in green text, or papers saved in her library folders rendered in red text (Figure~\ref{fig:types}). On the other hand, citations to unexplored papers are also prioritized with different shades of yellow highlights based on how many papers in her reading history also cited them (Figure~\ref{fig:types}). These personalized augmentations allow her to contextualize the current paper by surfacing citations to familiar papers and focusing her attention on examining the most important citations when trying to find other relevant papers to read. 

As she reads, she notices one of the inline citations about the training dataset used in the current paper was highlighted in a darker shade, indicating potential high relevance (Figure~\ref{fig:types}). However, she wonders if the training dataset is popular amongst image captioning researchers. To better make sense of the citation, she clicks on it to pop up a Paper Card (Figure~\ref{fig:paper_cards}) that contains more context(s) around the cited paper. She could see general information about the cited paper on the Paper Card, including the title and the abstract. She also notices that it was highly cited with 500 citations, suggesting that it is a high-quality and popular dataset. However, both the current and cited dataset papers were published more than six years ago, and she wonders if the dataset is still relevant today or if newer datasets have replaced it. To address this, she scrolls down in the Paper Card to find that two other papers from her recent readings have also cited the same dataset paper and were published this year. She looked at the citing sentences from both of these newer papers, confirming that it was still being used in recent work. Feeling more confident, she uses the bookmark button in the Paper Card to save the cited paper in her \emph{image captioning datasets} library folder.

As she continues to explore and read more papers, she notices the overview pages added by {\SYSTEM} to the top of each documents allow her to make quick judgements about the current paper (Figure~\ref{fig:hero}). For example, seeing that 6 out of the 45 citations in the current paper were already saved in her library gave her confidence that the current paper is relevant to her topic of interest; and seeing that she had previously opened 12 of the inline citations gives her a sense of overall progress that she had covered many important work in the area. As she explores and builds up more papers in her library and reading history, {\SYSTEM} continues to learn more about her interests, allowing her to prioritize unexplored inline citations most related to her interests and use their Paper Cards to access a persistent historical context to see how relevant past activities around each citation.


\subsection{[D1] Discover Relevant Citations}

\begin{figure}[t]
\centering
	\fbox{\includegraphics[width=0.84\linewidth]{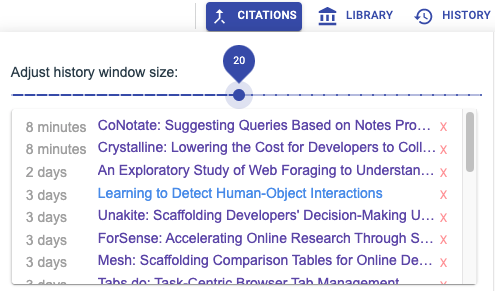}}
	\caption{Users can adjust the length of the inclusion history to include papers relevant to their current task.}
	\label{fig:list}
 \Description{The citation panel shows a horizontal slider, currently, 20 are selected. Below it there is a list of paper titles from the user's reading history. Left to the titles shows the last time each paper was opened (for example, 8 minutes or 2days)}
	\vspace{-3mm}
\end{figure}

Our first design goal was to help users more effectively explore inline citations during literature review by leveraging reencountered citations across papers in a user's reading history. For this, when opening a new paper in {\SYSTEM}, the PDF file is analyzed and linked to a paper entity on the Semantic Scholar academic graph using its APIs. This information allows {\SYSTEM} access to general information about both the current paper and papers listed in its references section, such as their titles, abstracts and citation counts. Details on how {\SYSTEM} processes raw PDFs of scientific papers are described in Section~\ref{sec:implementation-details}. After the current paper is analyzed, {\SYSTEM} compares citations in the current paper with citations in other papers the user had recently opened. Based on this comparison, the system then highlights the \textbf{reencountered citations} (Figure~\ref{fig:types}) that also appeared in multiple papers the user had recently opened, drawing the user's attention to them. To help users further prioritize their attention amongst reencountered citations, {\SYSTEM} uses different shades of yellow to orange highlighting based on the relevance of each inline citation (Figure~\ref{fig:types}). By default, {\SYSTEM} uses the 20 most recently read papers in the user's history, but this window can also be adjusted by the user using a slider to capture part of her reading history relevant to her current task (Figure~\ref{fig:list}).


{\SYSTEM} uses implicit and explicit signals that measure users' engagement levels with each paper to further scale the shades of the highlights.
We use the following heuristic for each inline citation to estimate its degree of interest for the current user: each paper in the reading history that cites the inline citation contributes 1 point. Then, an additional 0 to 1 point is added based on the estimated proportion it was read. For this, {\SYSTEM} tracks the maximum vertical scroll position proportional to the length of the paper to estimate the users’ reading progress. Users can also click on the \emph{read} button on the top menu bar to explicitly set this estimation to 100\% (Figure~\ref{fig:hero}). Two additional points are added for each citing paper saved in the library. For this, users can save a paper to their library using the bookmark icon in the menu bar (Figure~\ref{fig:hero}) or from a Paper Card (Figure~\ref{fig:paper_cards}). Finally, the total score is capped at 5 points when scaling the shades of the highlights. The assumption here is that opening a paper from the search results indicates moderate interest (1 point);  saving a paper to the library indicates a higher interest (2 points); and the more a paper is consumed, the higher the interest (0 to 1 points). 
This allows {\SYSTEM} to avoid over highlighting inline citations when a user opens many papers from a search result that might not all turn out to be relevant and important (a phenomenon we encountered in early design iterations.) Further, when a user opens a paper but later decides it is irrelevant, they  can also use the delete button in the menu bar (Figure~\ref{fig:hero}) to remove it entirely from their reading history, preventing it from contributing points to inline citations in other papers.

\subsection{[D2] Surfacing Familiar Papers}

Our first design goal aimed to direct the users' attention to inline citations to unexplored papers to help them discover and save important prior work relevant to their recent readings. In contrast, our second design goal aimed to surface citations to papers familiar to the current user, allowing them to contextualize the paper better. For this, {\SYSTEM} augments inline citations using two approaches. First, similar to how web browsers render visited and unvisited hyperlinks on web pages using different colors, citations to \textbf{visited papers} and \textbf{saved papers} in {\SYSTEM} are rendered in green and red, respectively (Figure~\ref{fig:types}).\footnote{In early design iterations we used blue and purple to be consistent with hyperlink in the browser, but we noticed a few publishers already use these colors for inline citations. Therefore, we switched to red and green to avoid confusion.}
This visual augmentation allows users to see which citations are to papers they had already explored or saved previously. In addition, when the user opens or saves a paper, its corresponding inline citation turns green or red in real-time, allowing her to keep track of which citations in the current paper were already covered. Second, {\SYSTEM} leverages a user's publication history as another source for finding papers familiar to the user, allowing the system to visually augment inline citations to the user's \textbf{own papers} and the papers cited by them (\textbf{cited papers}, Figures~\ref{fig:types} and \ref{fig:paper_cards}).

\subsection{[D3] Paper Cards with Personalized Context}

Making sense of inline citations can be challenging because the information in the current paper might be insufficient to judge their importance and relevance to the topic of interest. By highlighting inline citations in the current paper as described in the previous section, {\SYSTEM} allows users to identify reencountered citations amongst their recent readings more efficiently. However, users might need context beyond the citing sentence in the current paper to draw connections between a citation to different papers they had read. Our second design goal focused on providing personalized context around inline citations based on papers in a user's reading history to address this issue. When a user clicks on a highlighted reencountered citation, its Paper Card also contains the title of other papers in her history that contained the same citation. In the Paper Card, users can also examine the citing sentence across different papers, allowing them to see how the same paper was discussed across different papers they had read without manually cross-referencing between multiple documents. To further remind users how engaged they were with the citing papers, {\SYSTEM} also shows the last time each was opened and the estimated reading progress. Similarly Paper Cards of citations to familiar papers based on a user's publication history also show the titles of the user's publication and the relevant citing sentences. Finally, when saving a citation from its Paper Card, {\SYSTEM} keeps track of the context it was discovered, i.e., the current paper and the citing sentence. This context is then added to its Paper Card whenever the user re-encounters the same citation across different papers and in their library, as a way to remind them why it was saved in the first place (Figure~\ref{fig:paper_cards}, right).

In sum, {\SYSTEM} augments inline citations based on the current user's reading and publication history, allowing them to pay closer attention to citations to relevant or familiar papers for context and discovery. While it can be challenging for users to remember all the papers they had previously explored, the Paper Cards become a consistent and personalized context around each paper accessible whenever the users encounter the same citations while reading.


\subsection{Implementation Details}
\label{sec:implementation-details}
The front-end of {\SYSTEM} was built on the open-source ScholarPhi codebase \cite{Head2021AugmentingSP}. Around 5,000 lines of TypeScript and ReactJS code were added, resulting in around 17,000 lines of code in the final system.
Since many users primarily read papers from online sources or local files, we implemented {\SYSTEM} as a Chrome-extension which allowed us to become the default PDF reader for both their browsers and operating systems to ensure participants continued to engage with the system throughout the field deployment study. The backend was implemented in around 1,000 lines of Python using Flask and PostgreSQL to track user data, such as reading histories and behavior logging for our field study. We also use Grobid \cite{lopez2009grobid} for parsing and extracting citations and references from raw PDFs of scientific papers. To process PDFs on the fly, we set up Grobid in the server mode, allowing the front-end to upload PDF files to the backend for analysis. The Grobid server analyzes PDFs using a pre-trained conditional random fields model to identify the bounding boxes of inline citations and resolve them to the corresponding titles in the references. Processing time depends on the length of the PDFs, and a paper with 10-15 pages typically takes 5-15 seconds to process. During processing, users can freely browse the PDF document in {\SYSTEM} before the augmentations appear. We use the Semantic Scholar APIs to access users' publication history, paper libraries, and metadata about papers, such as their titles, abstracts, and abstract summaries generated by \cite{Cachola2020TLDRES}. 

To ensure no sensitive user data is compromised, we only automatically processed PDFs hosted on known domains of scientific paper archives (e.g., ACM, IEEE, AAAI, ArXiv, and ACL). For PDFs not hosted on known domains (including local files), {\SYSTEM} prompts users for permission to upload and process the PDF file for analysis. 
While the backend caches the processing results for repeated access to the same papers, the cached data only contains the coordinates of inline citations and their Semantic Scholar paper IDs. The uploaded PDF files are discarded after processing, and only a SHA1 hash of each file is kept for indexing. Finally, to ensure {\SYSTEM} is stable enough for field deployment, the research team used the extension internally to identify bugs and usability issues and provided feedback to improve the design of the system during the five months of development. 

\section{Study 1: Discover Relevant Citations}

\begin{figure*}[t]
\centering
	\includegraphics[width=1\linewidth]{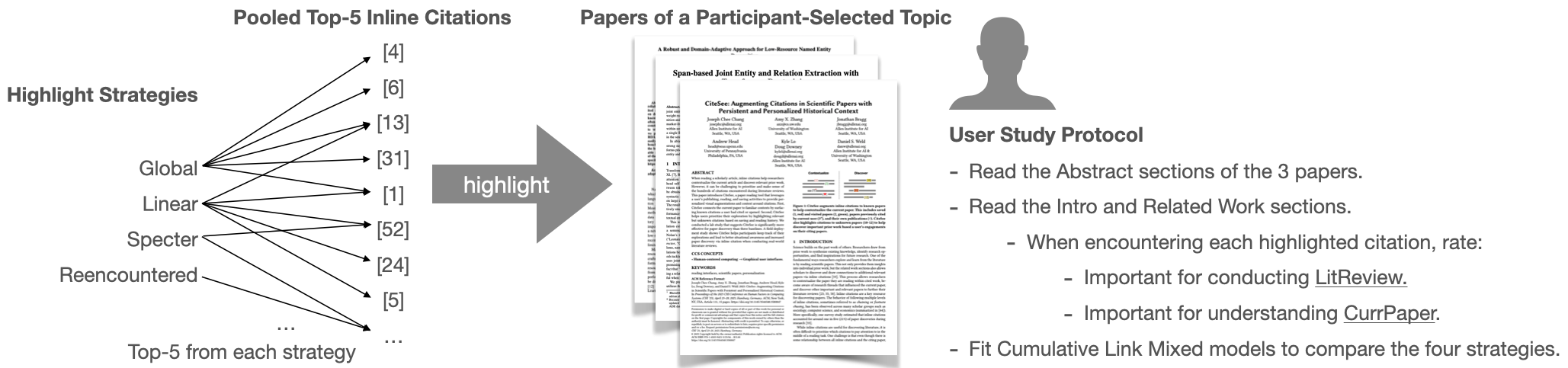}
	\caption{\new{Overview of Study 1. Comparing different highlight strategy for helping users to discover important inline citations for literature reviews. Participants rated all highlighted inline citation as each was encountered. Participants were blind to which strategies was used for each highlighted inline citations. In Study 2, we use the winning strategy as part of a field deployment study.}}
	\label{fig:study1_overview}
	\vspace{-3mm}
 \Description{ An overview figure of Study 1: On the left of the figure, it showed that there are four highlighting strategies for finding relevant inline citations. The figure showed each strategy linking to their top-5 inline citations. Different strategies can select the same inline citations. The selected citations are than pooled and used to highlight the 3 papers used in study 1 shown in the middle of the figure. On the right side of the figure, it showed a bullet point of the user study procedure at a high level. Here are the specific text shown in the figure: - Read the Abstract sections of the 3 papers. - Read the Intro and Related Work sections. - When encountering each highlighted citation, rate: --Important for conducting LitReview. -- Important for understanding CurrPaper.
- Fit Cumulative Link Mixed models to compare the four strategies.}
\end{figure*}

One of {\SYSTEM}'s core functionalities for supporting our first design goal is to highlight citations to relevant prior work during literature reviews. For this, {\SYSTEM} highlights citations in the current paper that are also cited by papers in a user's reading history. The main goal of Study 1 was to validate this approach by comparing it against three baselines using a controlled lab study. The study involved participants reading a set of three papers with the task of finding citations that were helpful in supporting a literature review scenario. \new{Figure~\ref{fig:study1_overview} shows the overview design of Study 1.}

To test in a more realistic literature review scenario, we collected real-world paper collections for three topics:

\begin{itemize}
\item \textbf{Topic 1}: Challenges UX designers face when working with unfamiliar AI technologies \cite{Yang2018InvestigatingHE,Dove2017UXDI,Dove2020MonstersMA}.
\item \textbf{Topic 2}: Top NLP techniques for extracting science concepts from research papers \cite{Eberts2020SpanbasedJE,Yu2020ARA,Beltagy2019SciBERTAP}.
\item \textbf{Topic 3}: Handheld controllers with haptic feedback for virtual reality applications \cite{Shigeyama2019TranscaliburAW,Chen2019HaptiVecPH,Sun2019PaCaPaAH}.
\end{itemize}

\noindent Specifically, Topic 1 was the required readings from a graduate-level HCI course,\footnote{\url{https://docs.google.com/document/d/1tR7G1ghLYpcqFj3v_E3CEBTAINIJuMTOumR1CCfykoo}} Topic 2 was the top-performing papers on the SciERC dataset \cite{Luan2018MultiTaskIO} public leaderboard\footnote{\url{https://paperswithcode.com/dataset/scierc}}, and Topic 3 was virtual reality controller papers published at the SIGCHI'19 conference \cite{chi19}.

At the beginning of the study, participants choose one of the three topics that most interested them to ensure engagement with the study. During the study, participants examined the three papers from their chosen topic in two passes to simulate activity in a literature review. The first pass was designed to build up their reading history and to learn the common theme of the three papers. The second pass was designed as a literature review task where they actively examined citations to find important prior work while justifying their choices and the signals they paid attention to by thinking out loud. More specifically, in the first pass, participants read the abstracts of the three papers and wrote a short summary of the common theme. 
The summaries they wrote were then used as the topic of interest for their literature review task in the second pass. 

After completing the first pass, the system pooled citations selected from the introduction and related work sections using four different strategies. To help control for the length of the study, we selected the top five citations using each strategy (random when tied). The four strategies are as follows:

\begin{itemize}
\item \textbf{Reencountered} Citations (System): Our personalized approach that selected citations also cited by the other two papers.\dw{the name common does NOT make it sound like it is the \SYSTEM strategy.  
i think this name will confuse people. note also, i reordered the items to put naive ones first} \kl{thoughts on name ``co-cite''?} \jc{i think people use cocite for [x,y,z] tho?} \kl{how about ``Consensus''?} 
\item \textbf{Linear} Reading: A non-personalized baseline that assumes users read papers linearly and select the first five citations.
\item \textbf{Global} Citation Count: A non-personalized baseline where citations to the five most highly cited papers are selected.
\item \textbf{Specter} Similarity \cite{Cohan2020SpecterDR}: A strong personalized baseline that uses semantic embeddings to select five citations that are the most similar to the mean vector of the three papers.

\end{itemize}

\noindent Since different strategies can select the same citations, this resulted in 12 to 20 total highlighted citations per paper and a total of 126 citations over the 9 papers tested. Of the 126 citations, 35 were selected by two strategies, 6 were selected by three strategies, and 2 were selected by four strategies.

At the start of the second pass, we walked through the different components in the Paper Cards, such as the global citation counts, citing sentences from the other two papers (when available), and abstract summaries (generated by \cite{Cachola2020TLDRES}) as described in the System section, participants could click on any citations as they read to see their corresponding Paper Cards. In addition, we also included the Specter embedding cosine similarity in the Paper Cards during Study 1 in all Paper Cards so that participants were exposed to the underlying signal for the Specter strategy. \new{During the second pass, participants were instructed to read through the Introduction and Related Work sections of each paper and to pay close attention to the highlighted citations. As they encountered each highlighted citation, participants answered the following two questions using a 5-point Likert scale for agreement in a separate survey form that listed the titles and reference numbers of highlighted citations and participants used in-page search to find the corresponding questions:}

\begin{itemize}
\item \textbf{Primary measure (LitReview)}: This paper is important for understanding the theme I wrote down. If I were to write a literature review, it would be important for me to read and include this paper.
\item \textbf{Secondary measure (CurrPaper)}: This paper is important for understanding the current paper.
\end{itemize}

\noindent Here, the first statement was the primary measurement of our study, and the second statement was designed to ensure participants were actively differentiating the topic of the current paper and the higher-level topic for their literature review scenario. \new{Finally, the study ended when a participant finished rating all of the highlighted inline citations.}

We recruited a total of ten participants across three universities and a research institute. To ensure participants were engaged in the tasks, we used convenience and snowball sampling to find participants who are likely to be interested in one or more of the topics we prepared (mean age: 28.3; SD:4.1; 8 male and 2 female; 8 PhD students, 1 postdoc, and 1 industry research scientist). Four of the participants chose Topic 1
, three chose Topic 2,
and three chose Topic 3.
The study took around 1 hour and each participant was compensated \$35 USD. This study was approved by our internal review board.

\subsection{Study 1 Limitations}

Literature reviews can be mentally taxing and time-consuming, making it challenging to study in a lab environment. \new{In early iterations of Study 1, we asked participants to deep read four papers while testing the four different highlighting strategies separately. While this design is more realistic and allowed participants to judge the four strategies individually and more holistically, it turned out to be too cognitively demanding and led to fatigue and failing to complete the study within 60-minutes.} We iteratively arrive at the final design of reading the first parts of three papers in two passes to simulate literature review activities. \new{This design was a compromise to control for the length and cognitive demand of the study, and it allowed us to compare multiple discovery highlighting strategy based on human judgements.}
While we only tested citations in the Introduction and Related Work sections, we believe citations are typically most concentrated in these sections. This design allowed us to ask participants to judge many citations while controlling for the amount of text they needed to read to maximize the data we can collect in 60-minute. Although participants were exposed to explanations related to the four strategies in the Paper Cards, we did not reveal how the citations were chosen. After the study, we revealed that they were rating citations highlighted by different strategies, and participants self-reported that they were not aware of it and assumed there was a single selection method. \new{To complement Study 1, we also conducted a field deployment in Study 2, where participants used {\SYSTEM} with the winning strategy we discovered in Study 1 for a prolonged period of time conducting their own real-world tasks.}

\subsection{Study 1 Results}
\label{sec:study-one-results}

Based on their think-aloud, participants engaged with the literature review scenarios and actively judged the connections between the cited papers and the summaries they generated in the first pass. For the same paper topics, participants generated similar summaries but with some variations for abstraction. For example, one participant who picked Topic 3 focused their literature review around \emph{hardware controllers that can simulate holding physical objects}, while another participant focused on \emph{techniques for depicting physical sensations}. Similarly, some participants who picked Topic 2 were focused on \emph{named entity recognition for scientific concepts} while others focused on the more general topic of \emph{named entity recognition for low-resources domains.} Participants also used a wide range of signals to make judgments about citations. Most immediately, they used the citing sentences and titles and abstracts of the cited paper to figure out how closely connected the citations were to the current paper and how relevant they were to the topic of interest. The two citation-based signals in the Paper Cards were also frequently mentioned. Specifically, this included the global citation counts and whether the citations were also cited by the other two papers in the assigned set. Most participants paid less attention to the Specter embedding distance on the Paper Cards. Instead, some mentioned that reading the titles and abstracts was often sufficient to see similarities between the three papers and cited papers, and the score sometimes acted as a validation of their judgment. 

\begin{figure}[t]
\centering
	\includegraphics[width=0.84\linewidth]{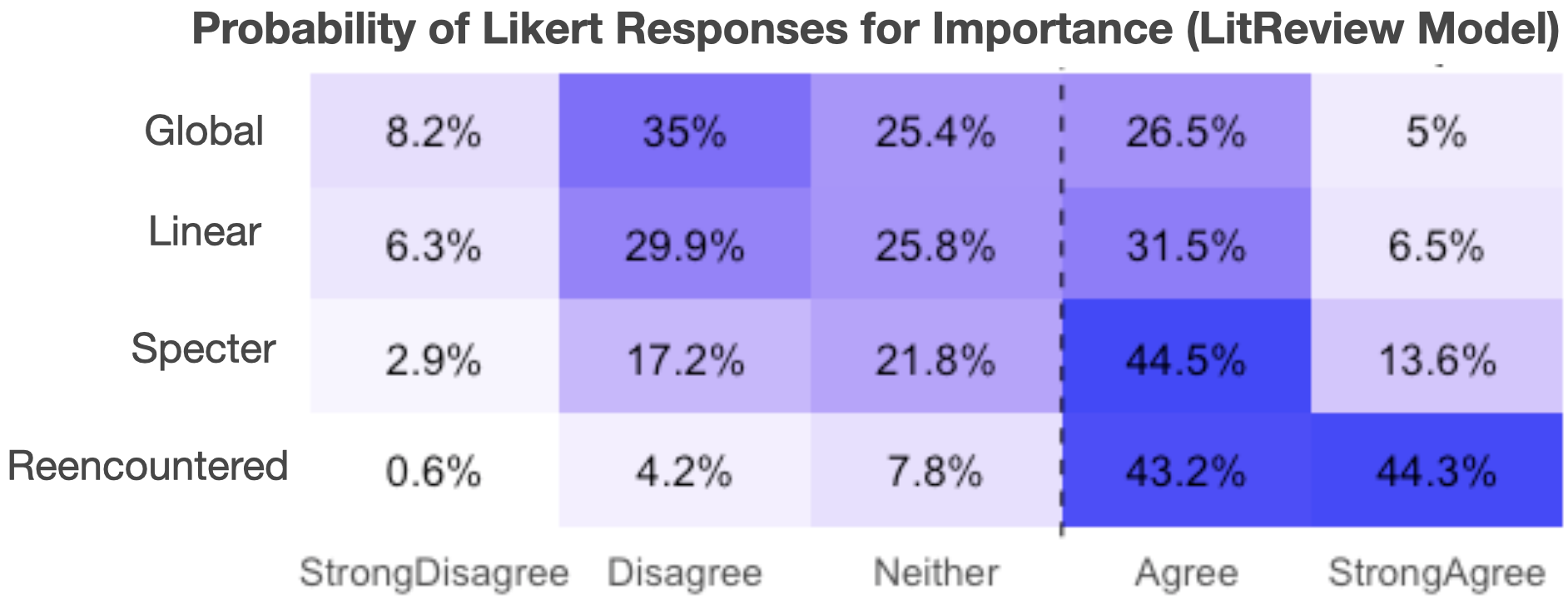}
	\caption{Probability of Likert responses of our reencountering highlighting strategies and three baselines in isolation based on a CLM model.}
	\label{fig:study1_probs}
 \Description{Probability of 5-point Likert responses of our reencountering highlighting strategies and three baselines in isolation based on a CLM model. The probability for each condition from Strongly Disagree to Strongly Agree are as follows. Global: 8.2\%, 35\%, 25.4\%, 26.5\%, 5\%; Linear: 6.3\%, 29.9\%, 25.8\%, 31.5\%, 6.5\%; Specter: 2.9\%, 17.2\%, 21.8\%, 44.5\%, 13.6\%; Reencountered: 0.6\%, 4.2\%, 7.8\%, 43.2\%, 44.3\%.}
	\vspace{-3mm}
\end{figure}

\begin{figure}[t]
\centering
	\includegraphics[width=1\linewidth]{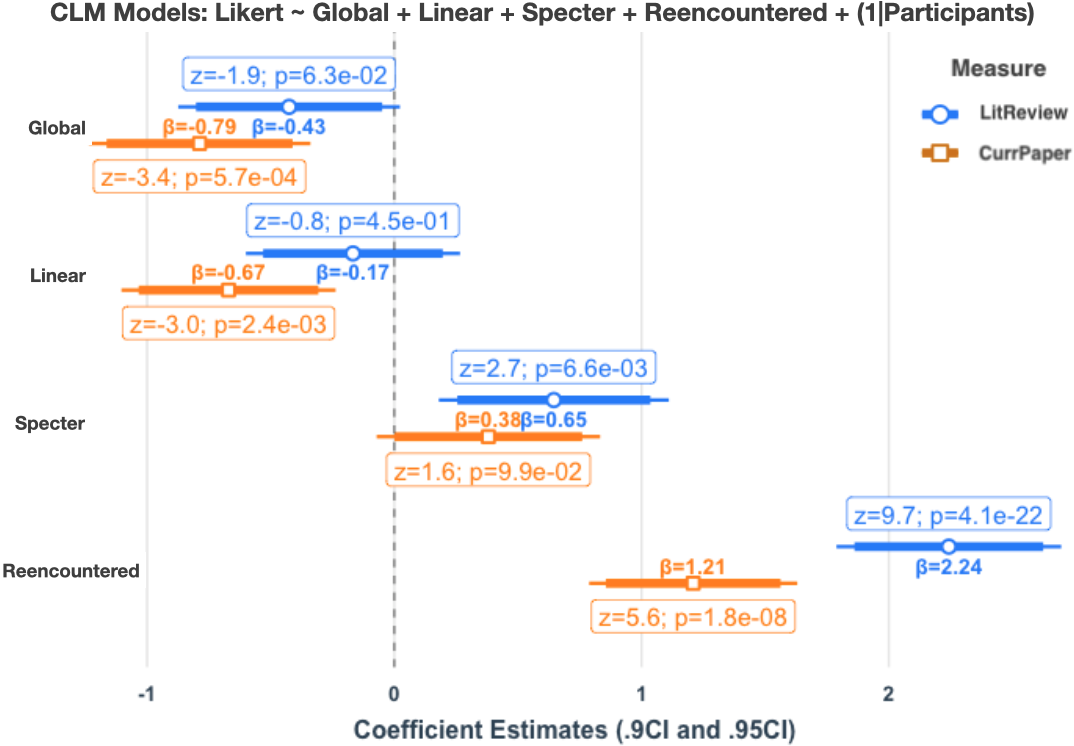}
	\caption{Study 1: Coefficients of four citation selection strategies based on an ordinal regression analysis for our primary (LitReview) and secondary (CurrPaper) measures. Confidence intervals that do not contain zero indicate significant correlation with the outcome Likert-scale ratings. A positive coefficient indicates the strategy is useful for supporting literature reviews (LitReview) or reading the current paper (CurrPaper). Combined with a permutation test, results suggest the system strategy (Reencountered) significantly outperformed the three baselines for both measures (Section~\ref{sec:study-one-results}).}
	\label{fig:study1}
 \Description{Confidence intervals of the coefficients from the LitReview and CurrPaper ordinal regression model. For the LitReview model, both coefficients for specter and common did not cross zero and were positive. For the CurrPaper model, the common coefficient was positive and did not cross zero. Both global and linear did not cross zero and were negative.}
	\vspace{-3mm}
\end{figure}

A total of 417 five-point Likert-scale responses were collected from the ten participants (180, 108, 129 responses for each topic, respectively).
We conducted a multiple regression analysis predicting the responses as a function of the fixed-effects of four strategies, represented as binary variables, and participant-specific random effects to account for variation in the user population and within-subjects correlation in responses. To handle ordinal Likert responses, we fit the following Cumulative Link Mixed (CLM) Model: 

    \vspace{3mm}
    {\small
    \noindent \textbf{Likert $\sim$ Reencountered + Specter + Global + Linear + (1\textbf{|}Participant)}
    }
    \vspace{0.1mm}

\noindent with a logit link using the R \emph{ordinal} package \cite{christensen2018cumulative};\footnote{\url{https://cran.r-project.org/web/packages/ordinal/index.html}} estimates for the fixed-effects $\beta$ of each strategy are reported in Figure~\ref{fig:study1}.\footnote{CLM models rely on an assumption of proportional odds---that is, each strategy has a similar effect on different levels of the Likert response. We verify this assumption is not violated via a Brant test \cite{brant-test} using the \emph{brant} library in R.} We fit one model for each of our primary and secondary measures---LitReview and CurrPaper. Figure~\ref{fig:study1} reports the estimated coefficients of different strategies and their confidence intervals, and Figure~\ref{fig:study1_probs} reports the estimated probabilities of Likert ratings for each condition in isolation.

We performed tests to verify the quality of fit of our CLM models to the observed data. First, to test whether \emph{any} strategy has a significant effect on the ratings, we performed a likelihood ratio test of our CLM model against a reduced model with only an intercept and subject-specific random effects.\footnote{\url{https://www.rdocumentation.org/packages/car/versions/3.0-12/topics/Anova}} For both measures, our full CLM model provides a significantly better fit to our data than the reduced model (LitReview: $\chi^2_{4}=151, p=2.2\text{e}{-16} < 0.001\text{***}$; CurrPaper: $\chi^2_{4}=88.967, p=2.2\text{e}-16 < 0.001\text{***}$).
Next, we considered inclusion of an extra fixed effect to control for paper topics, and found that the paper topics do not explain a significant amount of variability in the ratings that was not already captured in the strategy coefficients (LitReview $\chi^2_{2}=1.2777, p=0.5279$; CurrPaper $\chi^2_{2}=1.5998, p=0.4494$).  


Focusing on our primary measure LitReview results from Figure~\ref{fig:study1}, two strategies had a significant and positive effect on the Likert responses (Reencountered: $\beta=2.23, p=3.3\text{e}-22<0.001\text{***}$; Specter: $\beta=0.65, p=5.8\text{e}-3<0.01\text{**}$) while the other two baselines did not (Global: $\beta=-0.43, p=0.12$; Linear: $\beta=-0.27, p=0.30$). This result suggests that strategies that leverage reading history to find important citations for literature reviews outperform those that do not consider this personalized signal. To further compare the system strategy (Reencountered) and the Specter baseline, we performed a randomization test where we shuffled the strategy assignments in our data between Reencountered and Specter 1,000 times, fitting the same CLM model and recording the difference between Reencountered and Specter coefficients in each resulting model.
We found that the actual observed difference between the two coefficients significantly differed from the simulated differences, which fluctuated around zero, allowing us to conclude a significant difference in their effects ($\beta_{\text{reencountered}} - \beta_{\text{specter}} = 1.59,  p=3.91\text{e-}8<0.001\text{***})$. 
Taken together, our results suggest that the system approach (Reencountered) significantly outperforms the three baseline strategies.

\section{Study 2: Field Deployment}

We conducted a field deployment by recruiting participants from Study 1 who had planned to conduct a literature review within two weeks to further understand the costs and benefit of {\SYSTEM} in the real-world. Six participants were recruited (age: 36, 32, 27, 25, 33, 23; five male and one female). Each participant installed {\SYSTEM} on their personal and work computers for one to two weeks. Before the study, we briefly walked through all the features of {\SYSTEM} in 10 minutes. We also asked participants to keep a diary of their usage during the deployment via a feedback button in the system. We explicitly asked them to record interesting experiences using the reader and record at least six entries during the deployment.

Finally, we scheduled each participant for a 60-minute semi-structured post-interview one to two weeks after deployment based on their availability. During the post-interview, participants shared their screens and performed a retrospective walk-through of their experiences using {\SYSTEM}. This process included reopening papers they had read during the deployment and talking through their diary entries with us. All six participants completed the study and were each compensated \$35 USD for their time. The interviews were recorded and transcribed for an open thematic analysis to capture rich qualitative insights from their real-world usages \cite{Boyatzis1998TransformingQI,Connelly2013GroundedT}. \new{In our analysis below, even though we listed numbers of participants associated with each theme, we want to emphasize that we agree with  \citet{clarke2013successful} and \citet{vitale2018hoarding} that ``frequency does not determine value'' and that the goal of our deployment and interviews were not to capture distribution but to understand more deeply how {\SYSTEM}'s features can be used in real-world literature review tasks and the costs and benefits of adopting our inline citation augmentation approach.}
The study was approved by our internal review board.

\begin{table}[t]
  \centering
  \small

  \begin{tabular}{ l p{7.4cm}}
  
	\hline
	
	\textbf{ID} & \textbf{Literature Review Topics} \\
	
	\hline
	
	P1 & Co-design methodologies and examples. Interaction techniques for accessing video content.\\
	P2 & Tools for collecting textual information. Online sensemaking for programmers. HCI and large language models\\
	P3 & Interactive AutoML systems.\\
	P4 & Faceted retrieval interfaces of text documents.\\
	P5 & AI-supported qualitative data analysis.\\
	P6 & Solving math problems with neurosymbolic AI.\\
	
	\hline
	
  \end{tabular}
  \caption{Literature review topics conducted by participants in the field deployment study. P1 and P2 conducted multiple literature reviews and the rest focused on a single topic.}
  \Description{The literature review topics of different participants. P1: Co-design methodologies and examples. Interaction techniques for accessing video content. P2: Tools for collecting textual information. Online sensemaking for programmers. HCI and large language models. P3: Interactive AutoML systems. P4: Faceted retrieval interfaces of text documents. P5: AI-supported qualitative data analysis. P6: Solving math problems with neurosymbolic AI.}
  \label{tab:topics}
\end{table}

\subsection{Study 2 Results}

During post-interviews, participants retroactively walked through how they used {\SYSTEM} to conduct literature reviews. We found participants conducted a wide range of literature review topics, from design research to end-user interfaces and interaction techniques to machine learning (Table~\ref{tab:topics}). In particular, P1 and P2 each conducted three literature reviews on different topics, while P3-P6 focused on a single topic. 
\update{During deployment, {\SYSTEM} processed and augmented inline citations in the papers participants had opened. On average, each paper contained 41.0 (SD=24.8) inline citations. As expected, the majority of inline citations were not augmented (88.6\%, SD=24.5\%).  The augmented portion included 3.3\% (SD=8.1\%) of inline citations to papers familiar to the user (visited, saved, cited, and own papers, Figure~\ref{fig:types}), and 8.2\% (SD=9.6\%) that were also cited by papers in the user's reading history (reencountered citations, Figure~\ref{fig:types}). Based on the behavior logs, participants were also actively engaged with the system. We were initially concerned that highlighting the inline citations may be distracting to users, so we included a design where users could turn off the highlights and see relevant citations in a list view (Figure~\ref{fig:hero}). However, behavior logs showed that participants were actively engaged with augmented inline citations, and qualitative interviews showed seeing citations in-context helped users make connections across papers.}

\update{During the one-week deployment, each participant opened an average of 39.3 papers (SD=21) using {\SYSTEM} and saved 25.8 papers (SD=12.2) to their paper library (Table~\ref{tab:behavior}). Participants used both the inline citations and external sources (such as search engines or social recommendations) to explore and save useful papers. While a prior survey found inline citations accounted for around 21\% (n=881) of paper discovery during research \cite{King2009ScholarlyJI}, participants in our field deployment study used {\SYSTEM} to discover useful prior work via inline citations around 2.7x more frequently than that. On average, the majority of papers saved in our study came from examining inline citations (57\%, n=6, SD=24\%, .95CI[32\%,82\%]. Table~\ref{tab:behavior}) and the rest came from sources outside of the system that we did not track due to privacy concerns (e.g., web searches or social sharing.)}

\begin{table}[t]
\small
\centering
\begin{tabular}{lrrrrrr | r r}
\hline                          
\textbf{Action} & \textbf{P1} & \textbf{P2} & \textbf{P3} & \textbf{P4} &\textbf{P5} & \textbf{P6} & \textbf{AVG} & \textbf{SD}\\
\hline                          
Paper Opens &
77 &
48 &
17 &
33 &
33 &
28 &
39.3 &
21.0 \\

\hline

Card Opens &
62 &
96 &
155 &
207 &
74 &
71 &
110.8 &
57.9 \\

- FamiliarCite &
13{\scriptsize\%} &
15{\scriptsize\%} &
17{\scriptsize\%} &
4{\scriptsize\%} &
31{\scriptsize\%} &
35{\scriptsize\%} &
19{\scriptsize\%} &
12{\scriptsize\%} \\

- RencontrdCite &
42{\scriptsize\%} &
40{\scriptsize\%} &
49{\scriptsize\%} &
35{\scriptsize\%} &
20{\scriptsize\%} &
38{\scriptsize\%} &
37{\scriptsize\%} &
10{\scriptsize\%} \\

- NoAgmntation &
45{\scriptsize\%} &
46{\scriptsize\%} &
34{\scriptsize\%} &
60{\scriptsize\%} &
49{\scriptsize\%} &
27{\scriptsize\%} &
43{\scriptsize\%} &
12{\scriptsize\%} \\

\hline

Paper Saves &
33 &
14 &
29 &
14 &
45 &
20 &
25.8 &
12.2 \\

- FamiliarCite &
6{\scriptsize\%} &
14{\scriptsize\%} &
10{\scriptsize\%} &
7{\scriptsize\%} &
24{\scriptsize\%} &
10{\scriptsize\%} &
12{\scriptsize\%} &
7{\scriptsize\%} \\

- RencontrdCite &
9{\scriptsize\%} &
21{\scriptsize\%} &
34{\scriptsize\%} &
36{\scriptsize\%} &
44{\scriptsize\%} &
50{\scriptsize\%} &
33{\scriptsize\%} &
15{\scriptsize\%} \\

- NoAgmntation &
3{\scriptsize\%} &
14{\scriptsize\%} &
3{\scriptsize\%} &
29{\scriptsize\%} &
20{\scriptsize\%} &
5{\scriptsize\%} &
12{\scriptsize\%} &
10{\scriptsize\%} \\

- Search/External &
82{\scriptsize\%} &
50{\scriptsize\%} &
51{\scriptsize\%} &
29{\scriptsize\%} &
11{\scriptsize\%} &
35{\scriptsize\%} &
43{\scriptsize\%} &
24{\scriptsize\%} \\

\hline                          
\end{tabular}
  \caption{\update{Usage statistics from the field study (Study 2). Participants were actively engaged with the system during the deployment. They also used the Paper Cards in {\SYSTEM} to examine inline citations for both unexplored reencountered citations (M=37\%, SD=10\%), unexplored new citations (M=43\%, SD=12\%) and familiar papers (M=19\%, SD=12\%).}}
  \label{tab:behavior}
  \Description{Usage statistics of each participant. On average, each participant read 39.3 (SD=21.0) papers and saved 23.0 (SD=7.3) papers. They also frequently utilized the Paper Cards to access personalized context around inline citations. On average, each participant opened Paper Cards 110.8times (SD=57.9, Table 2). More than half (56\%) of the Paper Cards accessed contained personalized context. Specifically, around one in five Paper Cards accessed were to papers familiar to the user(M=19\%, SD=12\%). Common citations accounted for 37\% (SD=10\%).}
\vspace{-9.5mm}
\end{table}




\update{One explanation is that the inline citation augmentation provided by {\SYSTEM} improved the efficiency of discovering relevant prior work via inline citations. Evidently, participants were actively engaged with the reencountered citations highlighted by {\SYSTEM}, and used them to discover and save important prior work. For example, while reencountered citations only accounted for 8.2\% (SD=9.6\%) of the inline citations on average, they made up a disproportionate fraction of the Paper Cards accessed by the participants (mean=37\%, SD=10\%; or 4.5 times), suggesting that participants prioritized examining the reencountered citations. More importantly, beyond attracting users' attention to interact with the highlighted reencountered citations, we also found evidence that when participants examined reencountered citations, they had a near three times higher chance of discovering useful prior work. Specifically, while participants examined a similar number of augmented inline citations as they did unaugmented (43\% no augmentations vs 37\% reencountered, Table~\ref{tab:behavior}), reencountered citations accounted for nearly three times the number of papers saved from opening a Paper Card (M = 12\% vs 33\%).}

To better understand qualitative insights from the interviews, the first author went through the six hours of recording and transcripts in three passes to iteratively highlight interesting quotes and generate potential patterns until clear higher-level themes emerged \cite{Boyatzis1998TransformingQI,Connelly2013GroundedT}.
Overall, participants found value in using {\SYSTEM} for literature reviews. While some participants were initially overwhelmed trying to memorize the five different visual augmentation types, they gradually found more value after familiarizing themselves with {\SYSTEM} throughout the week. As expected, participants described a cold-start issue where the system initially only augmented few citations because their reading history was empty. However, as participants continued to read and save more papers, {\SYSTEM} was able to capture more signals about their literature review topics and convey them using different visual augmentations (Figure~\ref{fig:types}). Below we list the most common themes from the qualitative analysis to provide more insights into how these actions benefited participants during their literature reviews.

\subsubsection{Global Citation Counts vs Reencountered Citations (D1,D3)}\hfill



Since information about reencountered citations based on a user's reading history was new to the participants, we explicitly asked all participants to compare reencountered citations against the more familiar signal of global citation counts. Instead of ranking them, most participants found the two signals to be complementary and both useful. Specifically, most participants \new{(P1, P2, P3, P4, P5)} continued to see global citation counts as a proxy for estimating the quality for a cited paper, but found reencountered citations to be a better signal for judging relevance:


\begin{quote}
``They kind of serve different purposes. These [global citation] numbers are like \emph{``this paper alone, is it worth reading?''} And the stuff below [titles and citing sentences from other papers] actually showed me \emph{``is it related? and how is it related?}'' -- P2
\end{quote}




\noindent In particular, P3 described changing her prioritization of the two signals throughout the week. Specifically, as she built up her paper library and reading history, {\SYSTEM} was able to provide more value highlighting the reencountered citations:

\begin{quote}
``I remember when I first used [the system]...I looked at this [global citation counts] quite a lot...But as I got along, opening papers and added stuff to my library, more of these [reencountered citations] started appearing, \emph{now when I open the Paper Cards, I relied more on this [reencountered citations], rather than this [global citations}].'' -- P3
\end{quote}

\noindent One exception here is P6 who were cautious about global citation counts in general and the number of years since published, and said that she mostly relied on reading the titles and abstracts to determine paper quality and relevance even before the study.

\subsubsection{Highlighting based on Engagement with other Papers (D1)}\hfill


{\SYSTEM} highlighted reencountered citations in different shades of yellow to indicate potential relevance based on a user's engagement level with the citing papers. As participants retrospectively walked through paper they had opened during the deployment, we asked them to explicitly compare pairs of arbitrary reencountered citations with darker and lighter shades to probe on their accuracy. Participants said the different shades of highlights reflected the relevance of different inline citations based on their own interests and contexts \new{(P1, P2, P3)}. One representative example was P1 who conducted a literature review around \emph{interacting with video content} and had developed a higher interest around the subtopic of \emph{systems and interaction techniques} over \emph{video accessibility}:

\begin{quote}
``Related paper [highlighted reencountered citations in general] could either be about videos [interactions] or accessibility.` I'm not that interested in accessibility. So I think these [referring to citing papers in a Paper Card about accessibility with lowered engagement] are very accurate signals to me.'' -- P1
\end{quote}

\noindent On the other hand, one participant pointed to an edge case of spending significant amount of time on a paper only to realize it was unimportant:

\begin{quote}
``For papers that are in a different field, I might spend a lot of time to understand them but then decided it's not actually what I want. So there's like a mixed signal [referring to scroll tracking]'' -- P2
\end{quote}

\noindent In our design, users could remove papers from the reading history, which could potentially address this issue. However, when asked about this feature, participants either did not want to lose their reading history (P2) or did not think it was worth the effort (P1).

\subsubsection{Triaging between and within Papers (D1,D2)}\hfill


When opening a paper, {\SYSTEM} inserts a summary page at the top that lists the total count of different augmentation types (Figure~\ref{fig:hero}). The original intention was to provide the user a quick overview and to remind them of the different augmentation types supported by {\SYSTEM}. In the interviews, we found participants also used this summary to make quick judgements about the relevance of new papers they had opened \new{(P1, P2, P4, P5, P6)}:



\begin{quote}
``I would have 10 tabs opened about faceted search. Seeing these numbers popping up, I would switch their horizontal positions. So triaging, ranking, which I should look at was really helpful.'' -- P4
\end{quote}

In addition to prioritizing papers to read, participants also pointed to how {\SYSTEM} can support skimming papers \cite{SchemaladenPA,Hillesund2010DigitalRS,Nicholas2010ResearchersEU} because the reencountered citation highlights were often concentrated in sections that were more relevant to their literature review topics:

\begin{quote}
``It gives me confidence in making decisions of  which parts of the paper that I should pay attention to... And I did find that information really useful.'' -- P4
\end{quote}

\noindent Similar responses were also observed in \cite{Rachatasumrit2021ForSenseAO} for highlighting regions in webpages similar to a user's previous web clips. Here, we found evidence that highlighting citations in research papers based on a user's reading history had a similar and positive effect.

\subsubsection{Remembering Papers from the Past (D2,D3)}\hfill

Participants also described using {\SYSTEM} to help them recognize and remember previously explored papers \new{(P2, P3, P4, P5, P6)}. For example, P5 conducted a more exhaustive search of citations in two ``central'' papers, and used the visual augmentations (visited and saved papers) to keep track of the progress:

\begin{quote}
    ``One thing I like was when I save something it turns into a different color. It allows me to identify which papers are already in my reading list [library], and which ones need to be considered...It's a time consuming and error prone task [when not using {\SYSTEM}]'' -- P5
\end{quote}

When saving a cited paper from its Paper Card, {\SYSTEM} also records the current paper and the citing sentence to help users remember why a paper was saved in the future (Figure~\ref{fig:paper_cards}). When we retrospectively walked through papers in users' library, we probed whether this information provided value or if the titles and abstracts of saved papers were sufficient. Participants pointed to examples where this additional context was useful:

\begin{quote}
    ``Usually abstracts and titles does not give me enough information about why I saved it, because a paper can talk about a lot of different things. For example this title is super vague, but here [the saved from citing sentence], it talks about `time spend.' It gives me a better sense of why.'' -- P5
\end{quote}

\noindent While outside the scope of this work, participants also pointed out that further customization of context could be useful, such as attaching notes or specific sentences from the current paper.


\subsubsection{Sensemaking across Multiple Papers (D1,D2,D3)}\hfill



Leveraging different features in {\SYSTEM}, participants described how it improved their process of exploring many papers for literature reviews compared to existing processes. For example, P3 compared paper discovery in {\SYSTEM} to a search engine, and found using {\SYSTEM} led to finding more relevant prior work:

\begin{quote}
``I think this extension changes my workflow in a good way. I can immediately get context for the citations and find relevant things to add to my library. If I just use `AutoML' to search on Google Scholar, there's going to be a lot of not relevant things that's purely of ML techniques. But yeah, using the reader I can bring up papers that are more [relevant] on the HCI side of things.'' -- P3
\end{quote}

Finally, participants pointed to how using {\SYSTEM} provided better situational awareness of the connections between many papers \new{(P3, P4, P5)}. P5 in particular described how this allowed him to quickly identify ``central'' papers important in the domain and discover different sub-domains based on seeing patterns in the citing sentences connecting different pockets of work:

\begin{quote}
    ``You are tracking which papers I have opened or read, citing sentences, and which papers cited which papers. It give me a sense of the citation network and what this space is about. I can see `Cody' seems like \emph{a recent and central paper}. It also showed me people are doing \emph{mix initiative} stuff, and \emph{optimizing features} for qualitative analysis and \emph{active learning}. I can see common papers that are closely related to each other.'' -- P5
\end{quote}

\subsubsection{Volunteered Continued Usage in the Wild}
\update{Finally, we also found that 4 out of the 6 participants continued to be actively engaged with {\SYSTEM} after the study had concluded for more than two months (62, 74, 112, 121 days at the time of writing). Considering how they volunteered to use our research prototype under no obligations nor rewards, we see this as a promising indication that tightly integrating personalized paper discovery support in a reading environment can provide continued value to our participants in real-world settings.}




\section{Limitations and Future Work}

\update{
While {\SYSTEM} used a slider that allowed users to adjust the length of reading history considered by the system (Figure~\ref{fig:list}),
a future direction is to provide better support for users interleaving reading for multiple tasks.
Potential solutions includes allowing users to explicitly create and specify current literature review context (e.g., creating topical projects or library folders) or automatically identify parts of a reading history relevant to the current paper. 
Participants in study 2 agreed that our simple weighting heuristics based on scroll positions and saving to library led to highlighting shades that reflected their personal interests in the wild. Primarily, users sometimes open many papers from a search result but only to quickly close many after skimming the abstracts to filter our ones that were not relevant enough. This simple heuristic allowed {\SYSTEM} to avoid over-highlighting inline citations based on papers loaded in background tabs or less relevant papers quickly closed by the users. \new{In addition, the simple 5-point heuristic for reencountered citations could become over-saturated over prolonged usage (depending on the sparsity of citations in an area of interests). We did account for this when designing {\SYSTEM} in two ways: 1) When a user opens or saves a reencountered citation, they become ``familiar'' and are no longer highlighted in yellow. 2) In the paper cards, users could explicitly click on ``remove highlight'' to prevent it from being highlighted in the future. While participants in the week-long deployment did not point to having too few or too many citations being highlighted, we did observe an increase in proportion of inline citations being slightly highlighted in the second half of the week (M=13.0\% SD=10.2\%). Future work could conduct longer deployment studies to develop more sophisticated approaches to track and analyze users' reading behaviors for predicting paper relevance more accurately. For example, we could solicit gold-standard personalized paper relevance ratings and correlate them to behavior traces such as reading time and mouse hovering patterns \cite{rzeszotarski2011instrumenting}. However, prior work has also shown even users could face high uncertainty when highlighting manually during sensemaking tasks \cite{chang2016supporting}, and it is not apparent that further improving the highlight shading accuracy would have significant benefit. 
}}


\new{While exploiting citations to recommend papers can be powerful and easy for the users to understand, this common approach \cite{Mackinlay1995AnOU,Huang2002AGR,Gori2006ResearchPR,Chau2011ApoloMS,Xia2016ScientificAR,Ponsard2016PaperQuestAV,He2019PaperPolesFA} can potentially introduce echo chamber effects. 
One way to mitigate this is to incorporate semantic similarity signals into the highlighting strategy so that the system is able to also highlight prior work that has not yet been cited by many papers. For example, we could incorporate the Specter baseline from Study 1, which also had a significant and positive effect on the Likert-scale rating, to help users further triage inline citations that were not highlighted based on citation signals. This approach has the potential of nudging users to explore semantically similar prior work not are not commonly cited by their reading histories. Another potential direction is to show paper recommendation in the margins based on what was cited in the current paper, similar to the design shown in \cite{Rachatasumrit2022CiteReadIL}. 
Finally, a more aggressive approach could be suggesting further search query terms based on a user's reading history to further their breath of paper exploration using techniques similar to \cite{palani2021conotate,palani2022interweave}. However, presenting paper and query suggestions while not disrupting a user's reading flow is likely an important challenge. 
}

\new{Finally, while our work focuses on the exploration and discovery aspects of literature review, many participants also pointed to the potential of supporting manual note-taking and synthesis across multiple papers. One promising direction is to generalize the idea of consistent Paper Cards in this work and support scientific concepts in papers for learning.
For example, a practitioner learning about machine learning could bring up \emph{Concept Cards} for \emph{language models} and \emph{transformer models} when mentioned in the current paper and see prior notes and relevant paragraphs gathered from papers she has recently read to keep track of important concepts used across literature. Recent work both in NLP on linking scientific concepts \cite{cattan2021scico} and in HCI on in-situ web clipping, organization and maintaining provenance \cite{kuznetsov2022fuse,han2022passages} could potentially lead to techniques for driving this novel interaction for scholarly research support.}

\section{Conclusion}


In this paper, we introduce {\SYSTEM}, a novel scientific paper reading tool that provides a personalized literature review experience. {\SYSTEM} leverages a user's prior research activities to augment inline citations in the current paper, which helps the user contextualize the current paper and explore prior work relevant to their literature reviews. Our designs were motivated by an exploratory interview study and combined ideas from prior work in intelligent reading interfaces that focused on non-personalized, single-document reading support and recommender systems that focused on non-reading paper discovery. Through a lab study, we found our strategy of using reading history to augment inline citations to be significantly more effective in helping users discover prior work compared to three baseline strategies. Through a week-long field deployment study, participants conducting real-world literature reviews valued the additional personalized context around inline citations provided by {\SYSTEM}, which allowed them to have better situational awareness when exploring many papers.

\begin{acks}
This project is supported by NSF Grant OIA-2033558 and ONR Grant N00014-21-1-2707. The authors thank Marti A. Hearst, Matt Latzke, Evie Cheng, and Cassidy Trier for the insightful discussions and feedback. We also thank the anonymous reviewers for their constructive feedback. Finally, this work would not have been possible without our pilot test and user study participants. %
\end{acks}

\bibliographystyle{ACM-Reference-Format}
\bibliography{acmart}

\end{document}